\documentclass[runningheads]{llncs}
\usepackage[english]{babel}
\usepackage{amsmath}
\usepackage{amsfonts}
\usepackage[ruled, vlined, nofillcomment]{algorithm2e}
\usepackage{subcaption}
\usepackage{xspace}
\usepackage{url}
\usepackage{icomma}
\usepackage{enumitem}
\usepackage{booktabs}
\usepackage{tikz}
\usepackage{pgfplots}
\usepackage[numbers,sort&compress]{natbib}

\newcommand{\ifcondense}{\iftrue}
\newenvironment{IEEEproof}{\begin{proof}}{\end{proof}}

\newcommand{\set}[1]{\left\{#1\right\}}

\newcommand{\fpr}[1]{\mathopen{}\left(#1\right)}

\newcommand{\abs}[1]{{\left|#1\right|}}

\newcommand{\np}{\textbf{NP}}
\newcommand{\poly}{\textbf{P}}
\newcommand{\apx}{\textbf{APX}}
\newcommand{\naturals}{\mathbb{N}}
\newcommand{\integers}{\mathbb{Z}}
\newcommand{\funcdef}[3]{{#1}:{#2} \to {#3}}

\DeclareRobustCommand{\dispfunc}[2]{%
  \ensuremath{%
  \ifthenelse{\equal{#2}{}}%
    {\mathit{#1}}%
    {\mathit{#1}\fpr{#2}}}}

\newcommand{\score}[1]{\dispfunc{q}{#1}}
\newcommand{\pen}[1]{\dispfunc{p}{#1}}

\newcommand{\bigO}[1]{\dispfunc{\mathcal{O}}{#1}}
\newcommand{\rmin}[1]{\dispfunc{r_{\mathit{min}}}{#1}}
\newcommand{\rmax}[1]{\dispfunc{r_{\mathit{max}}}{#1}}

\newcommand{\flux}[1]{\dispfunc{fluc}{#1}}

\newcommand{\fasprb}{\textsc{FAS}\xspace}

\newcommand{\agstaticprb}{\textsc{agony}\xspace}
\newcommand{\aggenprb}{\textsc{gen-agony}\xspace}
\newcommand{\agsmoothprb}{\textsc{fluc-agony}\xspace}

\newcommand{\agsegprb}{\textsc{seg-agony}\xspace}

\newcommand{\agstaticab}{\textsc{agony}\xspace}
\newcommand{\agsmoothab}{\textsc{fluc}\xspace}

\newcommand{\agsegab}{\textsc{seg}\xspace}

\newcommand{\ctorprb}{\textsc{change2ranks}\xspace}
\newcommand{\rtocprb}{\textsc{ranks2change}\xspace}

\newcommand{\satprb}{\textsc{3SAT}\xspace}

\newcommand{\dtname}[1]{\textsl{#1}}

\SetKwComment{tcpas}{\{}{\}}
\SetCommentSty{textnormal}
\SetArgSty{textnormal}
\SetKw{False}{false}
\SetKw{True}{true}
\SetKw{Null}{null}
\SetKwInOut{Output}{output}
\SetKwInOut{Input}{input}
\SetKw{AND}{and}
\SetKw{OR}{or}
\SetKw{Break}{break}

\pgfdeclarelayer{background}
\pgfdeclarelayer{foreground}
\pgfsetlayers{background,main,foreground}

\definecolor{yafaxiscolor}{rgb}{0.3, 0.3, 0.3}

\definecolor{yafcolor1}{rgb}{0.4, 0.165, 0.553}
\definecolor{yafcolor2}{rgb}{0.949, 0.482, 0.216}
\definecolor{yafcolor3}{rgb}{0.47, 0.549, 0.306}
\definecolor{yafcolor4}{rgb}{0.925, 0.165, 0.224}
\definecolor{yafcolor5}{rgb}{0.141, 0.345, 0.643}
\definecolor{yafcolor6}{rgb}{0.965, 0.933, 0.267}
\definecolor{yafcolor7}{rgb}{0.627, 0.118, 0.165}
\definecolor{yafcolor8}{rgb}{0.878, 0.475, 0.686}

\tikzstyle{exnode} = [inner sep = 1pt]
\tikzstyle{labnode} = [sloped, text = black, font = \scriptsize, inner sep = 1pt]
\tikzstyle{exedge} = [yafcolor5, draw, thick, >=latex, ->]
\tikzstyle{exedge2} = [yafcolor2, draw, thick, >=latex, ->]
\tikzstyle{exedge3} = [yafcolor3, draw, thick, >=latex, ->]

\newlength{\yafaxispad}
\newlength{\yaftlpad}
\newlength{\yaflabelpad}
\newlength{\yafaxiswidth}
\newlength{\yafticklen}
\makeatletter
\def\pgfplots@drawtickgridlines@INSTALLCLIP@onorientedsurf#1{}
\makeatother

\newcommand{\yafdrawxaxis}[2]{
	\pgfplotstransformcoordinatex{#1}\let\xmincoord=\pgfmathresult 
	\pgfplotstransformcoordinatex{#2}\let\xmaxcoord=\pgfmathresult 
	\pgfsetlinewidth{\yafaxiswidth} 
	\pgfsetcolor{yafaxiscolor}
	\pgfpathmoveto{\pgfpointadd{\pgfpointadd{\pgfplotspointrelaxisxy{0}{0}}{\pgfqpointxy{\xmincoord}{0}}}{\pgfqpoint{-0.5\yafaxiswidth}{\yafaxispad}}}
	\pgfpathlineto{\pgfpointadd{\pgfpointadd{\pgfplotspointrelaxisxy{0}{0}}{\pgfqpointxy{\xmaxcoord}{0}}}{\pgfqpoint{0.5\yafaxiswidth}{\yafaxispad}}}
	\pgfusepath{stroke}

}
\newcommand{\yafdrawyaxis}[2]{
	\pgfplotstransformcoordinatey{#1}\let\ymincoord=\pgfmathresult 
	\pgfplotstransformcoordinatey{#2}\let\ymaxcoord=\pgfmathresult 
	\pgfsetlinewidth{\yafaxiswidth} 
	\pgfsetcolor{yafaxiscolor}
	\pgfpathmoveto{\pgfpointadd{\pgfpointadd{\pgfplotspointrelaxisxy{0}{0}}{\pgfqpointxy{0}{\ymincoord}}}{\pgfqpoint{\yafaxispad}{-0.5\yafaxiswidth}}}
	\pgfpathlineto{\pgfpointadd{\pgfpointadd{\pgfplotspointrelaxisxy{0}{0}}{\pgfqpointxy{0}{\ymaxcoord}}}{\pgfqpoint{\yafaxispad}{0.5\yafaxiswidth}}}
	\pgfusepath{stroke}
}

\newcommand{\yafdrawaxis}[4]{\yafdrawxaxis{#1}{#2}\yafdrawyaxis{#3}{#4}}

\pgfplotscreateplotcyclelist{yaf}{%
{yafcolor1,mark options={scale=0.75},mark=o}, 
{yafcolor2,mark options={scale=0.75},mark=square},
{yafcolor3,mark options={scale=0.75},mark=triangle},
{yafcolor4,mark options={scale=0.75},mark=o},
{yafcolor5,mark options={scale=0.75},mark=o},
{yafcolor6,mark options={scale=0.75},mark=o},
{yafcolor7,mark options={scale=0.75},mark=o},
{yafcolor8,mark options={scale=0.75},mark=o}} 

\pgfplotsset{axis y line=left, axis x line=bottom,
	tick align=outside,
	compat = 1.3,
	tickwidth=\yafticklen,
	clip = false,
	every axis title shift = 0pt,
    x axis line style= {-, line width = 0pt, opacity = 0},
    y axis line style= {-, line width = 0pt, opacity = 0},
    x tick style= {line width = \yafaxiswidth, color=yafaxiscolor, yshift = \yafaxispad},
    y tick style= {line width = \yafaxiswidth, color=yafaxiscolor, xshift = \yafaxispad},
    x tick label style = {font=\scriptsize, yshift = \yaftlpad},
    y tick label style = {font=\scriptsize, xshift = \yaftlpad},
    every axis y label/.style = {at = {(ticklabel cs:0.5)}, rotate=90, anchor=center, font=\scriptsize, yshift = -\yaflabelpad},
    every axis x label/.style = {at = {(ticklabel cs:0.5)}, anchor=center, font=\scriptsize, yshift = \yaflabelpad},
    x tick label style = {font=\scriptsize, yshift = 1pt},
    grid = major,
    major grid style  = {dash pattern = on 1pt off 3 pt},
	every axis plot post/.append style= {line width=\yafaxiswidth} ,
	legend cell align = left,
	legend style = {inner sep = 1pt, cells = {font=\scriptsize}},
	legend image code/.code={%
		\draw[mark repeat=2,mark phase=2,#1] 
		plot coordinates { (0cm,0cm) (0.15cm,0cm) (0.3cm,0cm) };%
	} 
}

\begin{document}

\title{Dynamic hierarchies in temporal directed networks}

\author{Nikolaj Tatti\inst{1,2}}
\authorrunning{N. Tatti}
\institute{F-Secure, Helsinki, Finland \and
Aalto University, Espoo, Finland
\email{nikolaj.tatti@aalto.fi}}

\maketitle

\begin{abstract} 
The outcome of interactions in many real-world systems can be often explained
by a hierarchy between the participants.
Discovering hierarchy from a given directed network 
can be formulated as follows: partition vertices into levels 
such that, ideally, there are only forward edges, that is, edges from upper levels to lower levels.
In practice, the ideal case is impossible, so instead we minimize some
penalty function on the backward edges. One practical option for such a penalty
is agony, where the penalty depends on the severity of the violation.
In this paper we extend the definition of agony to temporal networks.  In this
setup we are given a directed network with time stamped edges,
and we allow the rank assignment to
vary over time.
We propose 2 strategies for controlling the variation of individual ranks.  In
our first variant, we penalize the fluctuation of the rankings over time by
adding a penalty directly to the optimization function.  
In our second variant we allow the rank change at most once.
We show that the first variant can be solved exactly in polynomial time
while  the second variant is \np-hard, and in fact inapproximable.  However, we
develop an iterative method, where we first fix the change point and
optimize the ranks, and then fix the ranks and optimize the change points, and
reiterate until convergence.
We show empirically that the algorithms are reasonably fast in practice, and
that the obtained rankings are sensible.
\end{abstract}




\section{Introduction}\label{sec:intro}

The outcome of interactions in many real-world systems can be often explained
by a hierarchy between the participants. Such rankings occur in diverse
domains, such as, hierarchies among athletes~\citep{elo1978rating},
animals~\citep{jameson:99:behaviour,roopnarine:13:reef}, social network
behaviour~\citep{DBLP:conf/cse/MaiyaB09}, and browsing
behaviour~\citep{DBLP:conf/icdm/MacchiaBGC13}.

Discovering a hierarchy in a directed network can be defined as follows: given a
directed graph $G = (V, E)$, find an integer $r(v)$, representing a rank of
$v$, for each vertex $v \in V$, such that ideally $r(u) < r(v)$ for each edge
$(u, v) \in E$. This is possible only if $G$ is a DAG, so in practice, we penalize
each edge with a penalty $q(r(u), r(v))$, and minimize the total penalty.
One practical choice for a penalty is \emph{agony}~\citep{gupte:11:agony,tatti:14:agony,tatti:15:hierarchies}, $q(r(u), r(v)) = \max(r(u) - r(v) + 1, 0)$.
If $r(u) < r(v)$, an ideal case, then the agony is 0. On the other hand, if $r(u) = r(v)$, then
we penalize the edge by 1, and the penalty increases as the edge becomes more 'backward'.
The major benefit of computing agony is that we can solve it in polynomial time~\citep{gupte:11:agony,tatti:15:hierarchies,tatti:14:agony}.

In this paper we extend the definition of agony to temporal networks:  
we are given a directed network with time stamped edges\footnote{An edge
may have several time stamps.} and the idea is to allow the rank assignment to
vary over time; in such a case, the penalty of an edge with a time stamp $t$  depends only on the
ranks of the adjacent vertices at time $t$.

We need to penalize or constrain the variation of the ranks, as otherwise the
optimization problem of discovering dynamic agony
reduces to computing the ranks over individual snapshots. In order to do
so, we consider 2 variants. In our first variant, we compute the fluctuation of
the rankings over time, and this fluctuation is added directly to the
optimization function, multiplied by a parameter $\lambda$.  
In our second variant we allow the rank to change at most once, essentially
dividing the time line of a single vertex into 2 segments.

We show that the first variant can be solved exactly in $\bigO{m^2\log m}$
time.  On the other hand, we show that the second variant is \np-hard, and in
fact inapproximable.  However, we develop a simple iterative method, where we
first fix the change points and optimize the ranks, and then fix the ranks and
optimize the change points, and reiterate until convergence. We show that the
resulting two subproblems can be solved exactly in $\bigO{m^2\log m}$ time.

We show empirically that, despite the pessimistic theoretical running times, the
algorithms are reasonably fast in practice: we are able to compute the rankings
for a graph with over $350\,000$ edges in 5 minutes.

The remainder of the paper is organized as follows. We introduce the notation
and formalize the problem in Section~\ref{sec:prel}. In
Section~\ref{sec:static} we review the technique for solving static agony,
and in Section~\ref{sec:easy} we will use this technique to solve the first two
variants of the dynamic agony. In Section~\ref{sec:seg}, we present the iterative
solution for the last variant. Related work is given in Section~\ref{sec:related}.
Section~\ref{sec:exp} is devoted to experimental evaluation, and we conclude the
paper with remarks in Section~\ref{sec:conclusions}. The proofs for non-trivial
theorems are given in Appendix in supplementary material.

\section{Preliminaries and problem definition}\label{sec:prel}


We begin with establishing preliminary notation, and then continue by defining
the main problem.

The main input to our problem is a \emph{weighted temporal directed graph}
which we will denote by $G = (V, E)$, where $V$ is the set of vertices
and $E$ is a set of tuples of form $e = (u, v, w, t)$, meaning an edge $e$ from $u$ to $v$
at time $t$ with a weight $w$. We allow multiple edges to have the same time stamp,
and we also allow two vertices $u$ and $v$ to have multiple edges.
If $w$ is not provided we assume that an edge has a weight
of 1.
To simplify the notation we will often write $w(e)$ to mean
the weight of an edge $e$.
Let $T$ be the set of all time stamps.

A \emph{rank assignment} $\funcdef{r}{V \times T}{\naturals}$ is a function
mapping a vertex and a time stamp to an integer; the value $r(u; t)$ represents 
the rank of a vertex $u$ at a time point $t$.

Our next step is to penalize backward edges in a ranking $r$.
In order to do so, consider an edge $e = (u, v, w, t)$. We define the penalty
as
\[
	\pen{e; r} = w\times\max(0, r(u; t) - r(v; t) + 1)\quad.
\]
This penalty is equal to $0$ whenever $r(v; t) > r(u; t)$, if $r(v; t) = r(u;
t)$, then the $\pen{e; r} = w$, and the penalty increases as the difference
$r(u; t) - r(v; t)$ increases. 

We are now ready to define the cost of a ranking.
\begin{definition}
Assume an input graph $G = (V, E)$ and a rank assignment $r$.
We define a score for $r$ to be
\[
	\score{r, G} = \sum_{e \in E} \pen{e; r}\quad.
\]
\end{definition}

\textbf{Static ranking:}
Before defining the main optimization problems, let us first consider the
optimization problem where we do \emph{not} allow the ranking to vary over
time.

\begin{problem}[\agstaticprb]
\label{prb:static}
Given a graph $G = (V, E)$, an integer $k$, find a ranking $r$ minimizing
	$\score{r, G}$,
such that $0 \leq r(v; t) \leq k - 1$ and $r(v; t) = r(v; s)$,  for every $v \in V$ and $t, s \in T$.
\end{problem}

Note that \agstaticprb does not use any temporal information, in fact,
the exact optimization problem can be defined on a graph where we have stripped
the edges of their time stamps. This problem can be solved exactly in polynomial time,
as demonstrated by~\citet{tatti:15:hierarchies}. We should also point out that $k$ is an optional parameter,
and the optimization problem makes sense even if we set $k = \infty$.

\textbf{Dynamic ranking:} We are now ready to define our main problems.
 The main idea here is to allow the rank assignment to \emph{vary}
over time. However, we should penalize or constrain the variation of a ranking.
Here, we consider 2 variants for imposing such a penalty.

In order to define the first variant, we need a concept of fluctuation,
which is the sum of differences between the consecutive ranks of a given vertex.
\begin{definition}
Let $r$ be a rank assignment. Assume that $T$, the set of all time stamps, is
ordered, $T = t_1, \ldots, t_\ell$. The \emph{fluctuation} of a rank for a single
vertex $u$ is defined as
\[
	\flux{u; r} = \sum_{i = 1}^{\ell - 1} \abs{r(u, t_{i + 1}) - r(u, t_{i})}\quad.
\]
\end{definition}
Note that if $r(u, t)$  is a constant for a fixed $u$, then $\flux{u; r} = 0$.
We can now define our first optimization problem.

\begin{problem}[\agsmoothprb]
\label{prb:smooth}
Given a graph $G = (V, E)$, an integer $k$, and a penalty parameter $\lambda$,
find a rank assignment $r$ minimizing
\[
	\score{r, G} + \lambda \sum_{v \in V} \flux{v; r},
\]
such that $0 \leq r(v; t) \leq k - 1$ for every $v \in V$ and $t \in T$.
\end{problem}

The parameter $\lambda$ controls how much emphasis we would like to put in
constraining $\flux{}$: If we set $\lambda = 0$, then the $\flux{}$ term is
completely ignored, and we allow the rank to vary freely as a function of time. 
In fact, solving \agsmoothprb reduces to taking snapshots of $G$
at each time stamp in $T$, and applying \agstaticprb to these snapshots individually.
On the other hand, if we set $\lambda$ to be a very large number, then this
forces $\flux{v; r} = 0$, that is the ranking is constant over time. This
reduces \agsmoothprb to the static ranking problem, \agstaticprb.

In our second variant, we limit
how many \emph{times} we allow the rank to change. More specifically, we allow
the rank to change only once. 

\begin{definition}
We say that a rank assignment $r$ is a \emph{rank segmentation} if each $u$ changes its
rank $r(u; t)$ at most once. That is, there are functions
$r_1(u)$, $r_2(u)$ and $\tau(v)$
such that
\[
	r(u; t) =
	\begin{cases}
		r_1(u), & t < \tau(u), \\
		r_2(u), & t \geq \tau(u)\quad.
	\end{cases}
\]
\end{definition}

This leads to the following optimization problem.

\begin{problem}[\agsegprb]
\label{prb:agseg}
Given a graph $G = (V, E)$ and an integer $k$,
find a \emph{rank segmentation} $r$ minimizing $\score{r; G}$
such that 
$0 \leq r(v; t) \leq k - 1$
for every $v \in V$ and $t \in T$.
\end{problem}

Note that the obvious extension of this problem is to allow rank to change $\ell$
times, where $\ell > 1$. However, in this paper we focus specifically on the $\ell
= 1$ case as this problem yields an intriguing algorithmic approach, given in 
Section~\ref{sec:seg}.

\section{Generalized static agony}\label{sec:static}
In order to solve the dynamic ranking problems, we need to consider a minor
extension of the static ranking problem. 

To that end, we define a \emph{static graph} $H = (W, A)$ to be the graph, where
$W$ is a set of vertices and $A$ is a collection of directed edges $(u, v, c,
b)$, where $u, v \in V$, $c$ is a positive---possibly infinite---weight, and
$b$ is an integer, negative or positive.

\begin{problem}[\aggenprb]
Given a static graph $H = (W, A)$ find a function $\funcdef{r}{W}{\integers}$
minimizing
\[
	\sum_{(u, v, c, b) \in A} \max (c \times (r(u) - r(v) + b), 0)\quad.
\]
\end{problem}

Note that $c$ in $(u, v, c, b)$ may be infinite.  This implies that if the
solution has a finite score, then $r(u) + b \leq r(v)$.\!\footnote{Here we
adopt $0 \times \infty = 0$, when dealing with the case $r(u) - r(v) + b = 0$.}

We can formulate the static ranking problem, \agstaticprb, as an instance of
\aggenprb: Assume a graph $G = (V, E)$, and a(n optional) cardinality
constraint $k$. Define a graph $H = (W, A)$ as follows. The vertex set $W$
consists of the vertices $V$ and two additional vertices $\alpha$ and $\omega$.
For each edge $(u, v, w, t) \in E$, add an edge $(u, v, c = w, b = 1)$ to $A$.
If there are multiple edges from $u$ to $v$, then we can group them and combine the weights.
This guarantees that the sum in \aggenprb corresponds exactly to the cost function in \agstaticprb.
If $k$ is given, then add edges
$(\alpha, u, c = \infty, b = 0)$ and $(u, \omega, c = \infty, b = 0)$ for each $u \in V$.
Finally, add $(\omega, \alpha, c = \infty, b = 1 - k)$. This guarantees that
the for the optimal solution we must have $r(\alpha) \leq r(u) \leq r(\omega) \leq r(\alpha) + k - 1$,
so now the ranking defined $r(u; t) = r(u) - r(\alpha)$ satisfies the constraints by \agstaticprb. 

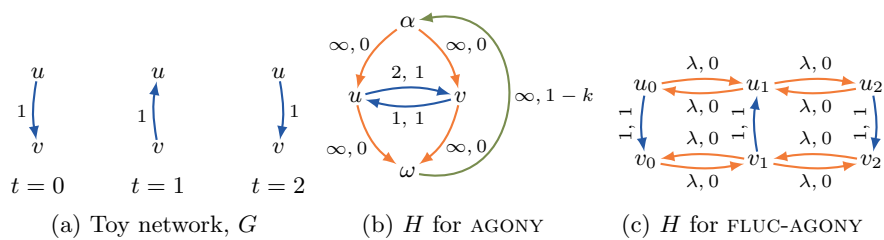
\begin{figure}
\hspace*{\fill}
\subcaptionbox{Toy network, $G$\label{fig:toy}}{
\begin{tikzpicture}

\node[exnode] at (0, 1) (p1) {$u$};
\node[exnode] at (0, 0) (n1) {$v$};
\node[exnode] at (0, -0.5)  {$t = 0$};

\draw (p1) edge[exedge, bend right = 10]  node[auto=right, black, font = \scriptsize, inner sep = 2pt] {1} (n1);

\begin{scope}[xshift = 1.6cm]
\node[exnode] at (-0, 1) (p2) {$u$};
\node[exnode] at (0, 0) (n2) {$v$};
\node[exnode] at (0, -0.5)  {$t = 1$};

\draw (n2) edge[exedge, bend left = 10]  node[auto, black, font = \scriptsize, inner sep = 2pt, pos = 0.4] {1} (p2);
\end{scope}

\begin{scope}[xshift = 3.2cm]
\node[exnode] at (0, 1) (p3) {$u$};
\node[exnode] at (0, 0) (n3) {$v$};
\node[exnode] at (0, -0.5)  {$t = 2$};

\draw (p3) edge[exedge, bend left = 10]  node[auto, black, font = \scriptsize, inner sep = 2pt] {1} (n3);
\end{scope}

\end{tikzpicture}}\hfill
\subcaptionbox{$H$ for \agstaticprb\label{fig:toystatic}}{
\begin{tikzpicture}
\node[exnode] at (-0.2, 0) (p1) {$u$};
\node[exnode] at (1.2, 0) (n1) {$v$};

\draw (p1) edge[exedge, bend left = 15]  node[auto, black, font = \scriptsize, inner sep = 2pt] {2, 1} (n1);
\draw (n1) edge[exedge, bend left = 15]  node[auto, black, font = \scriptsize, inner sep = 2pt] {1, 1} (p1);

\node[exnode] at (0.5, 1) (alpha) {$\alpha$};
\node[exnode] at (0.5, -1) (omega) {$\omega$};

\draw (alpha) edge[exedge2, bend right = 20] node[auto=right, pos = 0.5, black, font = \scriptsize, inner sep = 0pt] {$\infty, 0$} (p1);
\draw (alpha) edge[exedge2, bend left = 20] node[auto, black, pos = 0.5, font = \scriptsize, inner sep = 0pt] {$\infty, 0$} (n1);

\draw (p1) edge[exedge2, bend right = 20] node[auto=right, pos = 0.5, black, font = \scriptsize, inner sep = 1pt] {$\infty, 0$} (omega);
\draw (n1) edge[exedge2, bend left = 20] node[auto, black, pos = 0.5, font = \scriptsize, inner sep = 0pt] {$\infty, 0$} (omega);

\draw (omega) edge[exedge3, bend right = 100, looseness = 2] node[auto = right, black, font = \scriptsize, inner sep = 2pt] {$\infty, 1 - k$} (alpha);

\end{tikzpicture}}\hfill
\subcaptionbox{$H$ for \agsmoothprb\label{fig:toysmooth}}{
\begin{tikzpicture}

\node[exnode] at (0, 1) (p1) {$u_0$};
\node[exnode] at (0, 0) (n1) {$v_0$};

\draw (p1) edge[exedge, bend right = 10]  node[rotate = 90, anchor = south, pos = 0.5, black, font = \scriptsize, inner sep = 2pt] {1, 1} (n1);

\begin{scope}[xshift = 1.5cm]
\node[exnode] at (-0, 1) (p2) {$u_1$};
\node[exnode] at (0, 0) (n2) {$v_1$};

\draw (n2) edge[exedge, bend left = 10]  node[rotate = 90, anchor = south, pos = 0.5, black, font = \scriptsize, inner sep = 2pt] {1, 1} (p2);
\end{scope}

\begin{scope}[xshift = 3.0cm]
\node[exnode] at (0, 1) (p3) {$u_2$};
\node[exnode] at (0, 0) (n3) {$v_2$};

\draw (p3) edge[exedge, bend left = 10]  node[rotate = 90, anchor = south, pos = 0.5, black, font = \scriptsize, inner sep = 2pt] {1, 1} (n3);
\end{scope}

\draw (p1) edge[exedge2, bend left = 10] node[auto, black, font = \scriptsize, inner sep = 2pt] {$\lambda, 0$} (p2);
\draw (p2) edge[exedge2, bend left = 10] node[auto, black, font = \scriptsize, inner sep = 2pt] {$\lambda, 0$} (p3);

\draw (n1) edge[exedge2, bend right = 10] node[auto = right, black, font = \scriptsize, inner sep = 2pt] {$\lambda, 0$}  (n2);
\draw (n2) edge[exedge2, bend right = 10] node[auto = right, black, font = \scriptsize, inner sep = 2pt] {$\lambda, 0$}  (n3);

\draw (p3) edge[exedge2, bend left = 10] node[auto, black, font = \scriptsize, inner sep = 2pt] {$\lambda, 0$} (p2);
\draw (p2) edge[exedge2, bend left = 10] node[auto, black, font = \scriptsize, inner sep = 2pt] {$\lambda, 0$} (p1);

\draw (n3) edge[exedge2, bend right = 10] node[auto = right, black, font = \scriptsize, inner sep = 2pt] {$\lambda, 0$}  (n2);
\draw (n2) edge[exedge2, bend right = 10] node[auto = right, black, font = \scriptsize, inner sep = 2pt] {$\lambda, 0$}  (n1);

\end{tikzpicture}}
\hspace*{\fill}
\caption{Graph $G$, and the corresponding graphs $H$ used in \agstaticprb and \agsmoothprb.
In (b), the edges with omitted parameters have $c = \infty$ and $b = 0$. 
In (c), vertices $\alpha$ and $\omega$, and the adjacent edges, are omitted.}
\end{figure}

\begin{example}
\label{ex:toystatic}
Consider a temporal network given in Figure~\ref{fig:toy}.
The corresponding graph $H$ is given in Figure~\ref{fig:toystatic}.
\end{example}

As argued by~\citet{tatti:15:hierarchies}, \aggenprb is a dual problem of
capacitated circulation, a classic variant of a max-flow optimization problem.
This problem can be solved using an algorithm by~\citet{orlin:93:flow} in
$\bigO{\abs{A}^2\log \abs{W}}$ time.  In practice, the running time is faster.

\section{Solving fluc-agony}\label{sec:easy}

In this section we provide a polynomial solution for \agsmoothprb
by mapping the problem to an instance of \aggenprb.

Assume that we are given a temporal graph $G =
(V, E)$, a parameter $\lambda$ and a(n optional) constraint on the number of levels, $k$.

We will create a static graph $H = (W, A)$ for which solving
\aggenprb is equivalent of solving \agsmoothprb for $G$.
First we define $W$:
for each vertex $v \in V$ and a time stamp $t \in T$ such that there is an edge adjacent to $v$ at time $t$,
add a vertex $v_t$ to $W$. Add also two vertices $\alpha$ and $\omega$.
The edges $A$ consists of three groups $A_1$, $A_2$ and $A_3$:
\begin{enumerate}[label = {(\emph{\roman*})}, parsep = 1mm, nolistsep, leftmargin=0cm, itemindent = 7mm] 
\item For each edge $e = (u, v, w, t) \in E$, add
an edge $(u_t, v_t, c = w, b = 1)$.

\item Let $v_t, v_s \in W$ such that $s > t$ and there is no $v_o \in W$
with $t < o < s$, that is $v_t$ and $v_s$ are 'consecutive' vertices corresponding to $v$.
Add an edge $(v_t, v_s, c = \lambda, b = 0)$, also
add an edge $(v_s, v_t, c = \lambda, b = 0)$.

\item Assume that $k$ is given.
Connect each vertex $u_t$ to $\omega$ with $b = 0$ and
weight $c = \infty$.
Connect $\alpha$ to each vertex $u_t$ with $b = 0$ and
weight $c = \infty$.
Connect $\omega$ to $\alpha$ with $b = 1 - k$ and $c = \infty$.
This essentially forces $r(\alpha) \leq r(u_t) \leq r(\omega) \leq r(\alpha) + k - 1$.

\end{enumerate}

\begin{example}
Consider a temporal graph in Figure~\ref{fig:toy}.
The corresponding graph, without $\alpha$ and $\omega$, is given
in Figure~\ref{fig:toysmooth}.
\end{example}

Let $r$ be the rank assignment for $H$ with a finite cost, and define a rank assignment for $G$,
$r'(v; t) = r(v_t)$. The penalty of edges in $A_1$ is equal to $\score{r', G}$
while the penalty of edges in $A_2$ is equal to $\lambda \sum_{v \in V}
\flux{v, r'}$.
The edges in $A_3$ force $r'$ to honor the constraint $k$, otherwise $\score{r, H} = \infty$.
This leads to the following proposition.

\begin{proposition}
Let $r$ be the solution of \aggenprb for $H$.
Then $r'(v; t) = r(v_t) - r(\alpha)$ solves \agsmoothprb for $G$.
\end{proposition}

We conclude with the running time analysis. Assume $G$ with $n$ vertices and $m$ edges.
A vertex $v_t \in W$ implies that there is
an edge $(u, v, w, t) \in E$. Thus, $\abs{W} \in \bigO{m}$.
Similarly, $\abs{A_1} + \abs{A_2} + \abs{A_3} \in \bigO{m}$.
Thus, solving \aggenprb for $H$ can be done in $\bigO{m^2 \log m}$ time.

\section{Computing seg-agony}\label{sec:seg}
In this section we focus on \agsegprb.
Unlike the previous problem, \agsegprb is very hard to solve (see Appendix for the proof). 

\begin{proposition}
\label{prop:segnp}
Discovering whether there is a rank segmentation with a 0 score
is an \np-complete problem.
\end{proposition}

This result not only states that the problem is hard to solve exactly but it is
also very hard to approximate: there is no polynomial-time algorithm with a multiplicative approximation
guarantee, unless $\np=\poly$.

\subsection{Iterative approach}

Since we cannot solve the problem exactly, we have to consider a heuristic approach.
Note that
the rank assignment of a single vertex is characterized by 3 values: a change point, the rank
before the change point, and the rank after the change point.
This leads to the following iterative algorithm:
(\emph{i})
fix a change point for each vertex, and find the optimal ranks before and after the change point,
(\emph{ii})
fix the ranks for each vertex, and find the optimal change point.
Repeat until convergence.

More formally, we need to solve the following two sub-problems iteratively.

\begin{problem}[\ctorprb]
Given a graph $G = (V, E)$ and
a function $\tau$ mapping a vertex to a time stamp,
find
$\funcdef{r_1}{V}{N}$ and $\funcdef{r_2}{V}{N}$
mapping a vertex to an integer,
such that the rank assignment $r$ defined as
\[
	r(v; t) = 
	\begin{cases}
		r_1(v), & t < \tau(v), \\
		r_2(v), & t \geq \tau(v)
	\end{cases}
\]
minimizes $\score{r; G}$.
\end{problem}

\begin{problem}[\rtocprb]
Given a graph $G = (V, E)$ and
two functions
$\funcdef{r_1}{V}{N}$ and
$\funcdef{r_2}{V}{N}$
mapping a vertex to an integer,
find a rank segmentation $r$
minimizing $\score{r; G}$
such that there is a function $\tau$
such that
\[
	r(v; t) = 
	\begin{cases}
		r_1(v), & t < \tau(v), \\
		r_2(v), & t \geq \tau(v)\quad.
	\end{cases}
\]
\end{problem}

Surprisingly, we can solve both sub-problems exactly as we see in the next two
subsections. This implies that during the iteration the score will always
decrease. 
We still need a starting point for our iteration. Here, we initialize the
change point of a vertex $v$ as the median time stamp of $v$.

\subsection{Solving \ctorprb}
We begin by solving the easier of the two sub-problems.

Assume that we are given a temporal network $G = (V, E)$ and a function
$\funcdef{\tau}{V}{T}$. We will map \ctorprb to \aggenprb.
In order to do so, we define a graph $H = (W, A)$.
The vertex set $W$ consists of two copies of $V$; for each vertex $v \in V$, we create
two vertices $v^1$ and $v^2$, we also add vertices $\alpha$ and $\omega$ to enforce the constraint $k$. 
For each edge $e = (u, v, w, t) \in E$, we introduce an
edge $(u^i, v^j, c = w, b = 1)$ to $A$, where
\[
	i =
	\begin{cases}
		1 & \text{ if } t < \tau(u), \\
		2 &  \text{ if } t \geq \tau(u), \\
	\end{cases}
	\quad\text{and}\quad
	j =
	\begin{cases}
		1 & \text{ if } t < \tau(v), \\
		2 &  \text{ if } t \geq \tau(v)\quad. \\
	\end{cases}
\]
Finally, like before, we add $(\alpha, v, c = \infty, b = 0)$, $(v, \omega, c = \infty, b = 0)$
and $(\omega, \alpha, c = \infty, b = 1 - k)$ to enforce the constraint $k$.

We will denote this graph by $G(\tau)$.

\begin{example}
Consider the toy graph given in Figure~\ref{fig:toy}.
Assume $\tau(u) = 1$ and $\tau(v) = 2$. The resulting graph $G(\tau)$
is given in Figure~\ref{fig:toyc2r}.
\end{example}

The following proposition shows that optimizing agony for $H$
is equivalent of solving \ctorprb. We omit the proof as it is trivial.

\begin{proposition}
Let $r$ be a ranking for $H$.
Define $r'$ as
\[
	r'(v; t) = 
	\begin{cases}
		r(v^1) - r(\alpha), & t < \tau(v), \\
		r(v^2) - r(\alpha), & t \geq \tau(v)\quad.
	\end{cases}
\]
Then $\score{r', G} = \score{r, H}$.
Reversely, given a ranking $r'$ satisfying conditions of \ctorprb,
define a ranking $r$ for $G$ by setting $r(v^i) = r_i(v)$.
Then $\score{r', G} = \score{r, H}$.
\end{proposition}

We conclude with the running time analysis.
Assume $G$ with $n$ vertices and $m$ edges.
We have at most $2n + 2$ vertices in $W$ and
$\abs{A}  \in \bigO{m}$.
Thus, solving \ctorprb for $H$ can be done in $\bigO{m^2 \log n}$ time.

\subsection{Solving \rtocprb}
Our next step is to solve the opposite problem, where we are given the two alternative
ranks for each vertex, and we need to find the change points. Luckily, we can solve
this problem in polynomial time. To solve the problem we map it to \aggenprb, however unlike in previous
problems, the construction will be quite different.

Assume that we are given a graph $G = (V, E)$, and the two functions $r_1$ and
$r_2$. To simplify the following definitions, let us first define
\[
	\rmin{v} = \min(r_1(v), r_2(v)) \quad\text{and}\quad \rmax{v} = \max(r_1(v), r_2(v))\quad.
\]
Assume an edge $e = (u, v, w, t) \in E$. A solution to \rtocprb must
use ranks given by $r_1$ and $r_2$, that is
the rank of $u$ is either $\rmin{u}$ or $\rmax{u}$,
and the rank of $v$ is either $\rmin{v}$ or $\rmax{v}$,
depending where we mark the change point for $u$ and $v$.
This means that there are only 4 possible values for the penalty of $e$.
They are
\[
\begin{split}
	p_{00}(e) & = w\times\max(0, \rmin{u} - \rmin{v} + 1), \\
	p_{10}(e) & = w\times\max(0, \rmax{u} - \rmin{v} + 1), \\
	p_{01}(e) & = w\times\max(0, \rmin{u} - \rmax{v} + 1), \\
	p_{11}(e) & = w\times\max(0, \rmax{u} - \rmax{v} + 1)\quad.
\end{split}
\]
Among these penalties, $p_{01}(e)$ is the smallest, and ideally we would pay
only $p_{01}(e)$ for each edge. This is rarely possible, so we need to design a
method that takes other penalties into account.

Next we define a static graph $H = (W, A)$ that will eventually solve  \rtocprb.
For each vertex $v \in V$ and a time stamp $t \in T$
such that there is an edge adjacent to $v$ at time $t$,
add a vertex $v_t$ to $W$. Add also two additional vertices $\alpha$ and $\omega$.
We will define the edges $A$ in groups.
The first two sets of edges in $A$ essentially force $r(u_t) = 0, 1$, and that the ranking is monotonic as a function of $t$.
Consequenty, there will be at most only one time stamp for each vertex $u$, where the ranking changes.
This will be the eventual change point for $u$. The edges are:

\begin{enumerate}[label = {(\emph{\roman*})}, parsep = 1mm, leftmargin = 0pt, itemindent = 7mm]
\item Connect each vertex $u_t$ to $\omega$ with $b = 0$ and
weight $c = \infty$.
Connect $\alpha$ to each vertex $u_t$ with $b = 0$ and
weight $c = \infty$.
Connect $\omega$ to $\alpha$ with $b = -1$ and $c = \infty$.
Connect $\alpha$ to $\omega$ with $b = 1$ and $c = \infty$.
This forces $r(\alpha) \leq r(u_t) \leq r(\omega) = r(\alpha) + 1$.

\item Let $v_t, v_s \in W$ such that $s > t$ and there is no $v_o \in W$
with $t < o < s$.
If $r_2(v) \geq r_1(v)$, then connect $v_t$ to $v_s$
with $b = 0$ and $c = \infty$. This forces $r(v_s) \geq r(v_t)$.
If $r_2(v) < r_1(v)$, then connect $v_s$ to $v_t$
with $b = 0$ and $c = \infty$. This forces $r(v_s) \leq r(v_t)$.
\end{enumerate}
For notational simplicity, let us assume that $r(\alpha) = 0$.
The idea is then that once we have obtained the ranking for $H$, we can define the
ranking for $G$ as
\[
	r'(v; t) = \rmin{v} + (\rmax{v} - \rmin{v})r(v_t)\quad.
\]
Our next step is to define the edges that correspond to the penalties
in the original graph. We will show later in Appendix that the agony of $r'$ is equal to
$P_1 + P_2 + P_3 + \mathit{const}$, where
\[
\begin{split}
	P_1 & = \sum_{v_t \mid r(v_t) = 0} \sum_{e = (u, v, w, t) \in E} p_{00}(e) - p_{01}(e), \\
	P_2 & = \sum_{u_t \mid r(u_t) = 1} \sum_{e = (u, v, w, t) \in E} p_{11}(e) - p_{01}(e), \\
	P_3 & = \sum_{e = (u, v, w, t) \in E \atop r(v_t) = 0, r(u_t) = 1} p_{10}(e) - p_{00}(e) - p_{11}(e) + p_{01}(e)\quad.
\end{split}
\]
Let us first define the edges that lead to these penalties .

\begin{enumerate}[label = {(\emph{\roman*})}, parsep = 1mm, leftmargin = 0pt, itemindent = 7mm]
\item Connect $\omega$ to each vertex $v_t$ with $b = 0$ and weight
\[
	c = \sum_{e = (u, v, w, t) \in E} p_{00}(e) - p_{01}(e)\quad.
\]
In the sum $v$ and $t$ are fixed, and correspond to $v_t$.
This edge penalizes vertices with $r(v_t) = 0$ with a weight of $c$.
Summing these penalties yields $P_1$. 

\item Connect each vertex $u_t$ to $\alpha$ with $b = 0$ and
weight 
\[
	c = \sum_{e = (u, v, w, t) \in E} p_{11}(e) - p_{01}(e)\quad.
\]
In the sum $u$ and $t$ are fixed, and correspond to $u_t$.
This edge penalizes vertices with $r(u_t) = 1$ with a weight of $c$.
Summing these penalties yields $P_2$.

\item
For each edge $e = (u, v, w, t) \in E$, connect $u_t$ and $v_t$ 
with $b = 0$ and 
\[
	c = p_{10}(e) - p_{00}(e) - p_{11}(e) + p_{01}(e)\quad.
\]
This edge penalizes cases when $r(u_t) = 1$ and $r(v_t) = 0$,
and constitute $P_3$.
\end{enumerate}

We will denote the resulting $H$ by $G(r_1, r_2)$.

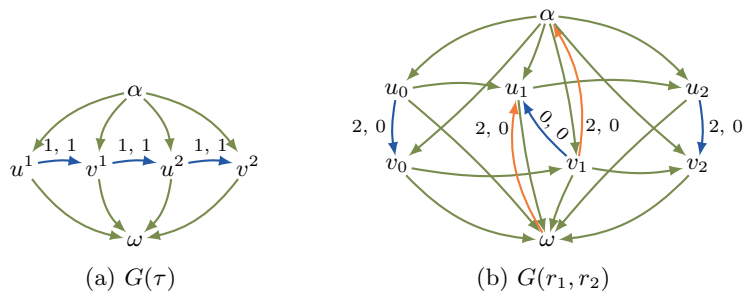
\begin{figure}
\hspace*{\fill}
\subcaptionbox{$G(\tau)$\label{fig:toyc2r}}{
\begin{tikzpicture}
\node[exnode] at (0, 0) (p1) {$u^1$};
\node[exnode] at (1, 0) (n1) {$v^1$};
\node[exnode] at (2, 0) (p2) {$u^2$};
\node[exnode] at (3, 0) (n2) {$v^2$};

\draw (p1) edge[exedge, bend left = 10]  node[auto, black, font = \scriptsize, inner sep = 2pt] {1, 1} (n1);
\draw (n1) edge[exedge, bend left = 10]  node[auto, black, font = \scriptsize, inner sep = 2pt] {1, 1} (p2);
\draw (p2) edge[exedge, bend left = 10]  node[auto, black, font = \scriptsize, inner sep = 2pt] {1, 1} (n2);

\node[exnode] at (1.5, 1) (alpha) {$\alpha$};
\node[exnode] at (1.5, -1) (omega) {$\omega$};

\draw (alpha) edge[exedge3, bend right = 20] (p1);
\draw (alpha) edge[exedge3, bend left = 20] (p2);
\draw (alpha) edge[exedge3, bend right = 20] (n1);
\draw (alpha) edge[exedge3, bend left = 20] (n2);

\draw (p1) edge[exedge3, bend right = 20] (omega);
\draw (p2) edge[exedge3, bend left = 20] (omega);
\draw (n1) edge[exedge3, bend right = 20] (omega);
\draw (n2) edge[exedge3, bend left = 20] (omega);

\end{tikzpicture}}\hfill
\subcaptionbox{$G(r_1, r_2)$\label{fig:toyr2c}}{
\begin{tikzpicture}

\node[exnode] at (0, 1) (p1) {$u_0$};
\node[exnode] at (0, 0) (n1) {$v_0$};

\draw (p1) edge[exedge, bend right = 10]  node[auto=right, black, font = \scriptsize, inner sep = 2pt] {2, 0} (n1);

\begin{scope}[xshift = 2cm]
\node[exnode] at (-0.4, 1) (p2) {$u_1$};
\node[exnode] at (0.4, 0) (n2) {$v_1$};

\draw (n2) edge[exedge, bend left = 10]  node[rotate = -45, pos = 0.7, auto=right, black, font = \scriptsize, inner sep = 0pt] {0, 0} (p2);
\end{scope}

\begin{scope}[xshift = 4cm]
\node[exnode] at (0, 1) (p3) {$u_2$};
\node[exnode] at (0, 0) (n3) {$v_2$};

\draw (p3) edge[exedge, bend left = 10]  node[auto, black, font = \scriptsize, inner sep = 2pt] {2, 0} (n3);
\end{scope}

\draw (p1) edge[exedge3, bend left = 10] (p2);
\draw (p2) edge[exedge3, bend left = 10] (p3);

\draw (n1) edge[exedge3, bend right = 10] (n2);
\draw (n2) edge[exedge3, bend right = 10] (n3);

\node[exnode] at (2, 2) (alpha) {$\alpha$};
\node[exnode] at (2, -1) (omega) {$\omega$};

\draw (n2) edge[exedge2, bend right = 20]  node[auto=right, pos = 0.2, black, font = \scriptsize, inner sep = 1pt] {2, 0} (alpha);

\draw (alpha) edge[exedge3, bend right = 20] (p1);
\draw (alpha) edge[exedge3, bend left = 5] (p2);
\draw (alpha) edge[exedge3, bend left = 20] (p3);

\draw (alpha) edge[exedge3, bend left = 5] (n1);
\draw (alpha) edge[exedge3, bend left = 5] (n2);
\draw (alpha) edge[exedge3, bend right = 5] (n3);

\draw (n1) edge[exedge3, bend right = 20] (omega);
\draw (n2) edge[exedge3, bend right = 5] (omega);
\draw (n3) edge[exedge3, bend left = 20] (omega);

\draw (p1) edge[exedge3, bend left = 5] (omega);
\draw (p2) edge[exedge3, bend right = 5] (omega);
\draw (p3) edge[exedge3, bend right = 5] (omega);

\draw (omega) edge[exedge2, bend left = 20]  node[auto, pos = 0.8, black, font = \scriptsize, inner sep = 1pt] {2, 0} (p2);

\end{tikzpicture}}
\hspace*{\fill}
\caption{
Graphs used for solving \agsegprb.
In both figures, the edges with omitted parameters have $c = \infty$ and $b = 0$.
For clarity, we omit edges between $\alpha$ and $\omega$ in both figures, in addition,
in (b) we omit parameters for the edges $(x, \alpha)$ and $(\omega, x)$ with $c = 0$.
}
\end{figure}

\begin{example}
Consider the toy graph given in Figure~\ref{fig:toy}.
Assume that the rank assignments are
$r_1(u) = 0$,
$r_1(v) = 1$,
$r_2(u) = 2$,
$r_2(v) = 3$.
The resulting graph $G(r_1, r_2)$ is given in Figure~\ref{fig:toyr2c}.
The optimal ranking for $G(r_1, r_2)$ assigns $0$ to
$\alpha$, $u_0$, $v_0$, and $v_1$; the rank for the remaining vertices is 1.
\end{example}

Before we show the connection between the ranks in $G$ and $H = G(r_1, r_2)$,
we first need to show that the edge weights are non-negative. This is needed to guarantee
that we can find the optimal ranking of $H$ using \aggenprb.

\begin{proposition}
The weights of edges in $H$ are non-negative.
\label{prop:segweightpositive}
\end{proposition}

The proof is given in Appendix.

We will state our main result: we can obtain the solution for \rtocprb
using the optimal ranking for $H$; see Appendix for the proof. 
\begin{proposition}
\label{prop:segtime}
Let $r$ be the optimal ranking for $H$.
Then
\[
	r'(v; t) = \rmin{v} + (\rmax{v} - \rmin{v}) (r(v_t) - r(\alpha))
\]
solves \rtocprb.
\end{proposition}

We conclude this section with the running time analysis.
Assume $G$ with $n$ vertices and $m$ edges.
A vertex $v_t \in W$ implies that there is
an edge $(u, v, w, t) \in E$. Thus, $\abs{W} \in \bigO{m}$.
Similarly, $\abs{A}  \in \bigO{m}$.
Thus, solving \rtocprb for $H$ can be done in $\bigO{m^2 \log m}$ time.

\section{Related work}\label{sec:related}

Perhaps the most classic way of ranking objects based on pair-wise interactions
is Elo rating proposed by~\citet{elo1978rating}, used to rank chess players.  A
similar approach was proposed by~\citet{jameson:99:behaviour} to model animal
dominance.

\citet{DBLP:conf/cse/MaiyaB09} proposed discovering
directed trees from weighted graphs such that parent vertices tend
to dominate the children. A hierarchy is evaluated by a
statistical model where the probability of an edge is high between a parent and
a child. A good hierarchy is then found by a greedy heuristic.

Penalizing edges using agony was first considered by~\citet{gupte:11:agony},
and a faster algorithm was proposed by~\citet{tatti:14:agony}. The setup was
further extended to handle the weighted edges, which was not possible with the
existing methods, by~\citet{tatti:15:hierarchies}, as well to be able to limit
the number of distinct ranks (parameter $k$ in the problem definitions). 

An alternative to agony is a penalty that penalizes an edge $(u, v)$ with $r(u)
\geq r(v)$ with a constant penalty. In such a case, optimizing the cost is
equal to \textsc{feedback arc set} (\fasprb), an \apx-hard problem with a
coefficient of $c = 1.3606$~\cite{dinur:05:cover}.  Moreover, there is no known
constant-ratio approximation algorithm for \fasprb, and the best known
approximation algorithm has ratio $O(\log n \log \log
n)$~\cite{even:98:feedback}.  In addition, \citet{tatti:15:hierarchies}
demonstrated that minimizing agony is \np-hard for any concave penalties while
remains polynomial for any convex penalty function. 

An interesting direction for future work is to study whether the rank obtained
from minimizing agony can be applied as a feature in role mining tasks, where
the goal is to cluster vertices based on similar
features~\citep{henderson:12:roix,mccallum:07:roles}.

\agsegprb essentially tries to detect a change point for each vertex. Change
point detection in general is a classic problem and has been studied
extensively, see excellent survey by~\citet{gama:14:survey}. However, these
techniques cannot be applied directly for solving \agsegprb since we would need to
have the ranks for individual time points.

The difficulty of solving \agsegprb stems from the fact that we allow vertices
to have different change points. If we require that the change point must be the equal
for all vertices, then the problem is polynomial. Moreover, we can easily
extend such a setup for having $\ell$ segments. Discovering change points then
becomes an instance of a classic segmentation problem which can be optimized
by a dynamic program~\citep{bellman:61:on}.

\section{Experiments}\label{sec:exp}

In this section we present our experimental evaluation.

\textbf{Datasets and setup:}
\label{sec:setup}
We considered 5 datasets. The first 3 datasets, \dtname{Mention},
\dtname{Retweet}, and \dtname{Reply}, obtained from SNAP
repository~\cite{snapnets}, are the twitter interaction networks related to
Higgs boson discovery. The 4th dataset, \dtname{Enron} consists of the email interactions
between the core members of Enron.
In addition, for illustrative purposes, we used a small dataset:
\dtname{NHL}, consisting of National Hockey League teams during the 2015--2016 regular season.
We created an edge $(x, y)$ if
team $x$ has scored more goals against team $y$ in a single game during the $2014$ regular season. We assign
the weight to be the difference between the points and the time stamp to be the date the game was played. 
We used hours as time stamps for Higgs
datasets, days for \dtname{Enron}. 
The sizes of the graphs are given in Table~\ref{tab:basic}.
\begin{table}[htb!]
\caption{Basic characteristics of the datasets and the experiments. The third data column, $\abs{T}$,
represents the number of unique time stamps, while the last column is the number of unique $(v, t)$ pairs
such that the vertex $v$ is adjacent to an edge at time $t$, $\big| \bigcup_{(u, v, w, t) \in E } \set{(v, t), (u, t)}\big|$.}

\label{tab:basic}
\begin{tabular*}{\columnwidth}{@{\extracolsep{\fill}}l rrrr}
\toprule

Name & $\abs{V}$ & $\abs{E}$ & $\abs{T}$ & $\abs{\set{(v, t)}}$ \\ 
\midrule
\dtname{Enron}   &
146 &
105\,522 &
964 &
24\,921 \\
\dtname{Reply} &
38\,683 &
36\,395 &
168 &
54\,892 \\
\dtname{Retweet}&
256\,491 &
354\,930 &
168 &
390\,583 \\
\dtname{Mention} &
115\,684 &
164\,156 &
168 &
183\,693 \\
\dtname{NHL}  &
30 &
1\,230 &
178 &
2\,460 \\
\bottomrule
\end{tabular*}
\end{table}

For each dataset we applied \agsmoothprb, \agsegprb, 
and the static variant, \agstaticprb.  
For \agsmoothprb we set $\lambda = 1$ for the Higgs datasets, $\lambda = 2$ for
\dtname{NHL} and \dtname{Enron}.

We implemented the algorithms in C++, and performed experiments using a
Linux-desktop equipped with a Opteron 2220 SE processor.\!\footnote{
See \url{https://bitbucket.org/orlyanalytics/temporalagony} for the code.}

\textbf{Computational complexity:}
First, we consider the running times, reported in
Table~\ref{tab:results}.  We see that even though the
theoretical running time is $\bigO{m^2 \log n}$ for \agsmoothprb
and for a single iteration of \agsegprb, the algorithms perform well in
practice.  We are able to process graphs with 300\,000 edges in 5 minutes.
Naturally, \agsegprb is the slowest as it requires multiple iterations---in our experiments 3--5 rounds---to converge.

\begin{table*}[htb!]

\caption{Agony, running time, and number of unique ranks in the ranking.}
\label{tab:results}
\setlength{\tabcolsep}{1pt}
\begin{tabular*}{\textwidth}{@{\extracolsep{\fill}}l rrr rrr rrr}
\toprule

&
\multicolumn{3}{l}{score} &
\multicolumn{3}{l}{number of ranks} &
\multicolumn{3}{l}{time}
\\
\cmidrule(r){2-4}
\cmidrule(r){5-7}
\cmidrule(r){8-10}
Name &
\agstaticab & \agsmoothab & \agsegab &
\agstaticab & \agsmoothab & \agsegab &
\agstaticab & \agsmoothab & \agsegab
\\
\midrule
\dtname{Enron}    &
57\,054 & 
21\,434 & 
50\,393 &

6 &
9 &
7 &

3s & 
4s &
26s 
\\

\dtname{Reply}    &     
6\,017 &
5\,401 &
4\,147 &

13 &
12 &
16 &

0.4s &
10s &
15s
\\

\dtname{Retweet}  &
2\,629 &
1\,384 &
1070 &

23 &
21&
18 &

8s &
4m &
5m
\\

\dtname{Mention}  &
12\,756 &
10\,082 &
8\,219 &

20 &
19 &
18 &

4s&
1m&
2m

\\
\dtname{NHL} &
2\,090 &
1\,414 &
1\,883 &

2 &
4 &
4 &

0.6s &
0.3s &
1s 
\\

\bottomrule
\end{tabular*}
\end{table*}

\begin{table*}[htb!]

\caption{Statistics measuring fluctuation of the resulting rankings:
$\flux{}$ is equal to the fluctuation $\flux{u; r}$ averaged over $u$,
$\mathit{maxdiff}$ is the maximum difference between the ranks of a single vertex $u$, averaged over $u$,
$\mathit{change}$ is the number of times rank is changed for a single vertex $u$, averaged over $u$.
Note that $\flux{} = \mathit{maxdiff}$ for \agsegprb as the assignment is allowed to change only once. 
}
\label{tab:results2}
\setlength{\tabcolsep}{1pt}
\begin{tabular*}{\textwidth}{@{\extracolsep{\fill}}l rr rr rr}
\toprule

&
\multicolumn{2}{l}{\flux{}} &
\multicolumn{2}{l}{$\mathit{maxdiff}$} &
\multicolumn{2}{l}{$\mathit{change}$}
\\
\cmidrule{2-3}
\cmidrule{4-5}
\cmidrule{6-7}
Name & 
\agsmoothab & \agsegab &
\agsmoothab & \agsegab &
\agsmoothab & \agsegab 
\\
\midrule
\dtname{Enron}    &

28.2 &
1 &

3.2 &
1 &

21.8 &
0.66 \\

\dtname{Reply}    &     

0.013 &
0.43 &

0.012 &
0.43 &

0.01 &
0.36 
\\
\dtname{Retweet}  &

0.003 &
0.17 &

0.003 &
0.17 &

0.002 &
0.13 \\

\dtname{Mention}  &

0.016 & 
0.3 &

0.014 &
0.3 &

0.011 &
0.2 

\\
\dtname{NHL} &

2.7 &
0.73 &

1.5 &
0.73 &

2.6 &
0.5

\\

\bottomrule
\end{tabular*}

\end{table*}

\textbf{Statistics of obtained rankings:}
Next, we look at the statistics of the obtained rankings, given in Table~\ref{tab:results}.  We
first observe that the agony of the dynamic variants is always lower than the
static agony, as expected.

Let us compare the constraint statistics, given in Table~\ref{tab:results2}.
First, we see that \agsmoothprb yields the smallest $\flux{}$ in Higgs databases.
\agsegprb produces smaller $\flux{}$ in the other two datasets but it also produces
a higher agony.

Interestingly enough, \agsmoothprb yields a surprisingly low
average number of change points for Higgs datasets. The low average is mainly due to most
resulting ranks being constant, and only a minority of vertices changing ranks over time.
However, this minority changes its rank more often than just once.

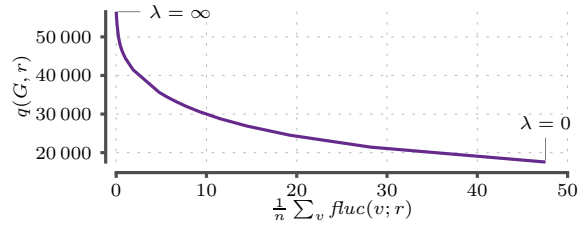
\begin{figure}
\begin{center}
\begin{tikzpicture}
\begin{axis}[xlabel={$\frac{1}{n}\sum_v \flux{v; r}$},ylabel= {$\score{G, r}$},
    width = 6cm,
    height = 2cm,
    cycle list name=yaf,
    scale only axis,
	tick scale binop=\times,
	x tick label style = {/pgf/number format/set thousands separator = {\,}},
	y tick label style = {/pgf/number format/set thousands separator = {\,}},
	scaled ticks = false,
	xmax = 50,
	xtick = {0, 10, 20, 30, 40, 50},
	pin distance = 3mm,
    no markers
    ]
\addplot table[x expr = {\thisrowno{2} / 146}, y index = 1, header = false] {enron_smooth_lambda.dat};
\node[black, inner sep = 0.5pt, pin={[black, font=\scriptsize]0:$\lambda = \infty$}] at (axis cs: 0, 56524) {};
\node[black, inner sep = 0.5pt, pin={[black, font=\scriptsize]90:$\lambda = 0$}] at (axis cs: 6943 / 146, 17566) {};
\pgfplotsextra{\yafdrawaxis{0}{50}{17566}{56524}}
\end{axis}
\end{tikzpicture}
\end{center}
\caption{Agony plotted against $\flux{}$ of the optimal ranking for \agsmoothprb by varying the parameter $\lambda$ (\dtname{Enron}).}
\label{fig:lambda_enron}
\end{figure}

\textbf{Agony vs fluctuation:}
The parameter $\lambda$ of \agsmoothprb provides a flexible way of controlling
the fluctuation: smaller values of $\lambda$ leads to smaller agony but larger
fluctuation while larger values of $\lambda$ leads to larger agony but smaller
fluctuation. This can be seen in Table~\ref{tab:results}, where relatively large $\lambda$
forces small fluctuation for the Higgs datasets, while relatively small $\lambda$
allows variation and a low agony for \dtname{Enron} dataset.
This flexibility comes at a cost: we need to have a sensible way
of selecting $\lambda$. One approach to select this value is to study the joint
behavior of the agony and the fluctuation as we vary $\lambda$.  This is
demonstrated in Figure~\ref{fig:lambda_enron} for \dtname{Enron} data, where we
scatter plot the agony versus the average fluctuation, and vary $\lambda$.  We see that
agony decreases steeply as we allow some fluctuation over time but the obtained
benefits decrease as we allow more variation. 

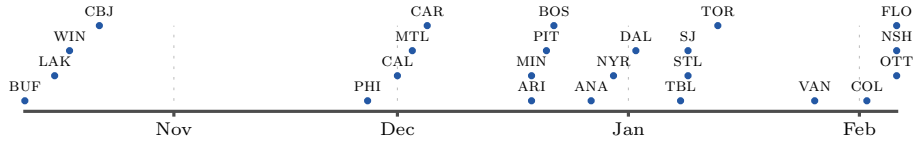
\begin{figure}[ht!]
\setlength{\tabcolsep}{3pt}
Ranking before the change:\\[1mm]
\begin{tabular*}{\columnwidth}{ll}
\toprule
1. &
\textsc{mtl win bos min dal flo wsh}\\
2. &
\textsc{van nyr chi sj tor col pit nsh stl tbl det nj nyi}\\
3. &
\textsc{cal lak ott buf car edm phi cbj ari ana} \\[2mm]
\bottomrule
\end{tabular*}\vspace{2mm}

Ranking after the change:\\[1mm]
\begin{tabular*}{\columnwidth}{ll}
\toprule
1. &
\textsc{nyr sj lak pit nsh stl tbl phi ana wsh} \\
2. &
\textsc{cal chi bos ott buf min dal car cbj det ari nj nyi flo} \\
3. &
\textsc{van mtl tor win col edm} \\
\bottomrule
\end{tabular*}\vspace{2mm}
Change points:\\[1mm]
\begin{tikzpicture}
\begin{axis}[
    width = 0.95\textwidth,
	height = 1cm,
    cycle list name=yaf,
    scale only axis,
	ytick = \empty,
	every node near coord/.append style = {font = {\scriptsize\sc}, text = black},
	xtick = {0, 31, 61, 92, 123, 152, 183},
	xticklabels = {Oct, Nov, Dec, Jan, Feb, Mar, Apr, May}
    ]
\addplot+[mark = *, mark size = 1pt, yafcolor5, nodes near coords, point meta = explicit symbolic, only marks]
	table[x index = 0, y index = 1, meta index = 2, header = false] {nhl_change.dat};
\pgfplotsextra{\yafdrawxaxis{11}{128}}
\end{axis}
\end{tikzpicture}
\caption{Rank segmentations for \dtname{NHL} with $k = 3$. The bottom figure shows only the teams whose rank changed. The $y$-axis is used only to reduce the clutter.}
\label{fig:nhl}
\end{figure}

\textbf{Use case:}
Finally, let us look on the rankings by \agsegprb of
\dtname{NHL} given in Figure~\ref{fig:nhl}. 
We limit the number of possible rank
levels to $k = 3$.

The results are sensible: the top teams are playoff teams while the
bottom teams have a significant losing record. Let us highlight some change points that reflect
significant changes in teams: for example, the collapse of \emph{Montreal Canadiens} (MTL)
from the top rank to the bottom rank coincides with the injury of their star goaltender.
Similarly, the rise of the \emph{Pittsburgh Penguins} (PIT) from the middle rank to the top rank
reflects firing of the head coach as well as retooling their strategy, \emph{Penguins} eventually won the Stanley Cup. 

\section{Concluding remarks}\label{sec:conclusions}

In this paper we propose a problem of discovering a dynamic hierarchy in a
directed temporal network. To that end, we propose two different optimization
problems: \agsmoothprb and \agsegprb. These problems vary in the
way we control the variation of the rank of single vertices. We show that
\agsmoothprb can be solved in polynomial time while \agsegprb
is \np-hard. We also developed an iterative heuristic for \agsegprb. 
Our experimental validation showed that the algorithms are practical, and the
obtained rankings are sensible.

\agsmoothprb is the more flexible of the two methods as the parameter
$\lambda$ allows user to smoothly control how much rank is allowed
to vary.  This comes at a price as the user is required to select an appropriate
$\lambda$.  One way to select $\lambda$ is to vary the parameter and monitor
the trade-off between the agony and the fluctuation. An interesting variant of
\agsmoothprb---and potential future line of work---is to minimize agony while
requiring that the fluctuation should not increase over some given threshold. 

The relation between \agsegprb and the sub-problems \rtocprb and \ctorprb is
intriguing: while the joint problem \agsegprb is \np-hard not only the
sub-problems are solvable in polynomial time, they are solved with the same
mechanism. 

A straightforward extension for \agsegprb is to allow more than just one change
point, that is, in such a case we are asked to partition the time line of each
vertex into $\ell$ segments. However, we can no longer apply the same iterative
algorithm. More specifically, the solver for \rtocprb relies on the fact that
we need to make only one change. Developing a solver that can handle the
more general case is an interesting direction for future work.

\bibliography{bibliography}

\newpage
\appendix
\section{Proofs of the propositions}

\subsection{Proof of Proposition~\lowercase{\ref{prop:segweightpositive}}}
\label{sec:proofsegweight}
To prove the proposition, we need the following lemma.
\begin{lemma}
\label{lem:diffsubmod}
Assume that we are given three numbers $a$, $b$, and $d$ with $a, b \geq 0$.
Define
\[
	h(x, y) = \max(ax - by + d, 0),
\quad\text{where}\quad x, y \in \set{0, 1}.
\]
Then $h(1, 1) - h(0, 1) \leq h(1, 0) - h(0, 0)$.
\end{lemma}

\begin{IEEEproof}
Straightforward calculation leads to
\[
	h(1, 1) - h(0, 1)  = \min(a, h(1, 1))
	\quad\text{and} \quad
	h(1, 0) - h(0, 0)  = \min(a, h(1, 0))
	\quad.
\]
Since $b \geq 0$, it follows that $h(1, 1) \leq h(1, 0)$, making the left equality smaller.
\qed
\end{IEEEproof}

\begin{IEEEproof}[of Proposition~\lowercase{\ref{prop:segweightpositive}}]
The inequality
$\rmax{u} \geq \rmin{u}$ implies $p_{11} - p_{01} \geq 0$, and so the weights of
the edges $(u_t, \alpha)$ are non-negative.  Similarly, $p_{00} - p_{01} \geq
0$, and so the weights of the edges $(\omega, u_t)$ are non-negative.

Assume edge $(u_t, v_t, c, 0)$, let
$a = \rmax{u} - \rmin{u}$,
$b = \rmax{v} - \rmin{v}$,
$d = 1 + \rmin{u} - \rmin{v}$. Since $a, b \geq 0$, Lemma~\ref{lem:diffsubmod}
states that $c  = p_{10}(u, v) - p_{00}(u, v) - p_{11}(u, v) + p_{01}(u, v) \geq 0$.
\qed
\end{IEEEproof}

\subsection{Proof of Proposition~\lowercase{\ref{prop:segtime}}}
\label{sec:proofsegtime}
Define $F$, a mapping transforming a ranking for $H$ to a ranking for $G$, as
\[
	F(r) = \rmin{v} + (\rmax{v} - \rmin{v}) (r(v_t) - r(\alpha))\quad.
\]

Let $R$ be the set of rankings for $H$ with $\score{r, H} < \infty$.
To prove the proposition, we show (\emph{i}) that the scores of $r \in R$
and $F(r)$ differ only by a constant, and (\emph{ii}) $\set{F(r) \mid r \in R}$
correspond to the valid rankings for \rtocprb.
These two results will immediately prove the Proposition~\ref{prop:segtime}.

\begin{lemma}
For $r \in R$,
	$\score{r, H} = \score{F(r), G} + \mathit{const}$.
\end{lemma}

\begin{IEEEproof}
Let $r \in R$. We can safely assume that $r(\alpha) = 0$.
Since edges $(u, v, c, b) \in A$ with $c = \infty$ guarantee
$r(v) \leq b + r(u)$, we have $0 = r(\alpha) \leq r(v_t) \leq
r(\omega) = 1$.

Let us split edges in $E$ in four groups $E_{00}$, $E_{01}$, $E_{10}$, $E_{11}$:
an edge $e = (u, v, w, t)$ belongs to $E_{xy}$ if $r(u_t) = x$ and $r(v_t) = y$.
Define $C_{xy} = \sum_{e \in E_{xy}} p_{xy}(e)$.
Then
\[
	\score{F(r), G} = C_{00} + C_{01} + C_{10} + C_{11}\quad.
\]

The cost $\score{r, H}$ consists of three parts.
The first part
is caused by the edges $(\omega, v_t)$ s.t. $r(v_t) = 0$, and it is equal to
\[
\begin{split}
	P_1 & = \sum_{v_t \mid r(v_t) = 0} \sum_{e = (u, v, w, t) \in E} p_{00}(u, v) - p_{01}(u, v) \\
	    & = \sum_{e \in E_{00} \cup E_{10}} p_{00}(e) - p_{01}(e)\quad.
\end{split}
\]
The second part
is caused by edges $(u_t, \alpha)$ for which $r(u_t) = 1$, and it is equal to
\[
\begin{split}
	P_2 & = \sum_{u_t \mid r(u_t) = 1} \sum_{e = (u, v, w, t) \in E} p_{11}(u, v) - p_{01}(u, v) \\
	    & = \sum_{e \in E_{11} \cup E_{10}} p_{11}(e) - p_{01}(e)\quad.
\end{split}
\]
The final part consists of edges between $u_t$ and $v_t$ for which $r(u_t) = 1$
and $r(v_t) = 0$, and it is equal to
\[
	P_3 = \sum_{e \in E_{10}} p_{10}(e) - p_{00}(e) - p_{11}(e) + p_{01}(e)\quad.
\]
Write $Z = \sum_{e \in E}  p_{01}(e)$.
Combining these leads us
\[
\begin{split}
	P_1 + P_2 + P_3 & = C_{00} + C_{11} + C_{10} -  \sum_{e \in E \setminus E_{01}}  p_{01}(e) \\
	 & = C_{00} + C_{11} + C_{10} + C_{01} - Z \\ 
	 & = \score{F(r), G} - Z, 
\end{split}
\]
where $Z$ does not depend on $r$.
\qed
\end{IEEEproof}

\begin{lemma}
$r'$ is a valid solution for \rtocprb
if and only if there is $r \in R$ with $r' = F(r)$.
\end{lemma}

\begin{IEEEproof}
Let $r \in R$, and let $r' = F(r)$.
We can safely assume that $r(\alpha) = 0$.
Since edges $(u, v, c, b) \in A$ with $c = \infty$ guarantee
$r(v) \leq b + r(u)$, we have $0 = r(\alpha) \leq r(v_t) \leq r(\omega) = 1$.
Consequently, $r'(v; t) = r_1(v)$ or $r'(v; t) = r_2(v)$.
Assume $r_1(v) < r_2(v)$. Then $r(v_t)$ is increasing, 
\[
	r(v_t) = 
	\begin{cases}
	1 & t \leq \tau \\
	0  & t > \tau
	\end{cases}
\]
for some $\tau$. Also, $\rmin{v} = r_1(v)$  and $\rmax{v} = r_2(v)$. So
\[
	r'(v; t) =
	\begin{cases}
	r_1(v) & t \leq \tau \\
	r_2(v)  & t > \tau\quad.
	\end{cases}
\]
The case $r_1(v) > r_2(v)$ is symmetric. Thus, $r'$ is a valid solution for \rtocprb.

Assume that you are given $r'$, a valid solution for Problem.
Define $r(v_t) = 1$ if $\rmax{v} = r'(v; t)$, and $0$ otherwise.
Extend the solution by setting $r(\alpha) = 0$ and $r(\omega) = 1$.
It is easy to see that for
any edge $(u, v, c, b) \in A$ with $c = \infty$ we have
$r(v) \leq b + r(u)$. Thus, $r \in R$.
\qed
\end{IEEEproof}

\subsection{Proof of Proposition~\lowercase{\ref{prop:segnp}}}
\label{sec:proofsegnp}

\begin{IEEEproof}
The problem is clearly in \np. We will use \satprb to prove the hardness.
Assume that we are given an instance of \satprb with $n$ variables and $m$ clauses.
We will prove the proposition in several steps.

\emph{Step 1 (graph construction):}
The graph consists of 3 vertex groups.
The first group consists of $2n$ vertices, $p_i$ and $n_i$, where $i = 1, \ldots, n$.
The second group consists of $3m$ vertices $C$, each vertex representing an occurrence of a literal in a clause.
For notational simplicity,
we will index vertices in $C$ by $c_{j\ell}$, where $j = 1, \ldots, m$ and $\ell = 1, 2, 3$.
The third group consists of $9m$ vertices $x_{j\ell}$, $y_{j\ell}$, and $z_{j\ell}$, where $j = 1, \ldots, m$ and $\ell = 1, 2, 3$.
In total, we have 3 unique time stamps, and the edges are given in Figure~\ref{fig:npproof}.

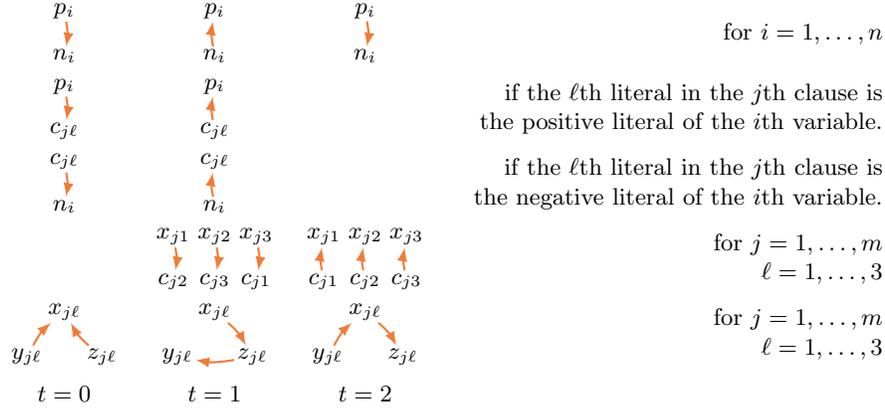
\begin{figure}
\begin{tikzpicture}

\node[exnode] at (0, 0.6) (p1) {$p_i$};
\node[exnode] at (0, 0) (n1) {$n_i$};

\draw (p1) edge[exedge2, bend left = 10] (n1);

\begin{scope}[xshift = 2cm]
\node[exnode] at (0, 0.6) (p2) {$p_i$};
\node[exnode] at (0, 0) (n2) {$n_i$};

\draw (n2) edge[exedge2, bend left = 10] (p2);
\end{scope}

\begin{scope}[xshift = 4cm]
\node[exnode] at (0, 0.6) (p3) {$p_i$};
\node[exnode] at (0, 0) (n3) {$n_i$};

\draw (p3) edge[exedge2, bend left = 10] (n3);

\end{scope}

\begin{scope}[xshift = 11cm]
\node[anchor = east] at (0, 0.3) {for $i = 1, \ldots, n$};
\end{scope}

\begin{scope}[xshift = 0cm, yshift = -1cm]
\node[exnode] at (0, 0.6) (a) {$p_i$};
\node[exnode] at (0, 0) (b) {$c_{j\ell}$};

\draw (a) edge[exedge2, bend left = 10] (b);
\end{scope}

\begin{scope}[xshift = 2cm, yshift = -1cm]
\node[exnode] at (0, 0.6) (a) {$p_i$};
\node[exnode] at (0, 0) (b) {$c_{j\ell}$};

\draw (b) edge[exedge2, bend left = 10] (a);
\end{scope}

\begin{scope}[xshift = 11cm, yshift = -1cm]
\node[align = right, anchor = east] at (0, 0.3) {if the $\ell$th literal in the $j$th clause is\\ the positive literal of the $i$th variable.};
\end{scope}

\begin{scope}[xshift = 0cm, yshift = -2cm]
\node[exnode] at (0, 0) (a) {$n_i$};
\node[exnode] at (0, 0.6) (b) {$c_{j\ell}$};

\draw (b) edge[exedge2, bend left = 10] (a);
\end{scope}

\begin{scope}[xshift = 2cm, yshift = -2cm]
\node[exnode] at (0, 0) (a) {$n_i$};
\node[exnode] at (0, 0.6) (b) {$c_{j\ell}$};

\draw (a) edge[exedge2, bend left = 10] (b);
\end{scope}

\begin{scope}[xshift = 11cm, yshift = -2cm]
\node[align = right, anchor = east] at (0, 0.3) {if the $\ell$th literal in the $j$th clause is\\the negative literal of the $i$th variable.};
\end{scope}

\begin{scope}[xshift = 2cm, yshift = -3cm]

\node[exnode] at (-0.55, 0.6) (b1) {$x_{j1}$};
\node[exnode] at (-0.55, 0) (d1) {$c_{j2}$};
\node[exnode] at (0, 0.6) (b2) {$x_{j2}$};
\node[exnode] at (0, 0) (d2) {$c_{j3}$};
\node[exnode] at (0.55, 0.6) (b3) {$x_{j3}$};
\node[exnode] at (0.55, 0) (d3) {$c_{j1}$};

\draw (b1) edge[exedge2, bend left = 10] (d1);
\draw (b2) edge[exedge2, bend left = 10] (d2);
\draw (b3) edge[exedge2, bend left = 10] (d3);
\end{scope}

\begin{scope}[xshift = 4cm, yshift = -3cm]
\node[exnode] at (-0.55, 0.6) (b1) {$x_{j1}$};
\node[exnode] at (-0.55, 0) (d1) {$c_{j1}$};
\node[exnode] at (0, 0.6) (b2) {$x_{j2}$};
\node[exnode] at (0, 0) (d2) {$c_{j2}$};
\node[exnode] at (0.55, 0.6) (b3) {$x_{j3}$};
\node[exnode] at (0.55, 0) (d3) {$c_{j3}$};

\draw (d1) edge[exedge2, bend left = 10] (b1);
\draw (d2) edge[exedge2, bend left = 10] (b2);
\draw (d3) edge[exedge2, bend left = 10] (b3);
\end{scope}

\begin{scope}[xshift = 11cm, yshift = -3cm]
\node[align = right, anchor = east] at (0, 0.3) {for $j = 1, \ldots, m$\\ $\ell = 1, \ldots, 3$};
\end{scope}

\begin{scope}[xshift = 0cm, yshift = -4cm]
\node[exnode] at (0, 0.6) (a) {$x_{j\ell}$};
\node[exnode] at (-0.5, 0) (b) {$y_{j\ell}$};
\node[exnode] at (0.5, 0) (c) {$z_{j\ell}$};
\node[exnode] at (0, -0.5)  {$t = 0$};

\draw (b) edge[exedge2, bend left = 10] (a);
\draw (c) edge[exedge2, bend left = 10] (a);
\end{scope}

\begin{scope}[xshift = 2cm, yshift = -4cm]
\node[exnode] at (0, 0.6) (a) {$x_{j\ell}$};
\node[exnode] at (-0.5, 0) (b) {$y_{j\ell}$};
\node[exnode] at (0.5, 0) (c) {$z_{j\ell}$};
\node[exnode] at (0, -0.5)  {$t = 1$};

\draw (c) edge[exedge2, bend left = 10] (b);
\draw (a) edge[exedge2, bend left = 10] (c);
\end{scope}

\begin{scope}[xshift = 4cm, yshift = -4cm]
\node[exnode] at (0, 0.6) (a) {$x_{j\ell}$};
\node[exnode] at (-0.5, 0) (b) {$y_{j\ell}$};
\node[exnode] at (0.5, 0) (c) {$z_{j\ell}$};
\node[exnode] at (0, -0.5)  {$t = 2$};

\draw (b) edge[exedge2, bend left = 10] (a);
\draw (a) edge[exedge2, bend left = 10] (c);
\end{scope}

\begin{scope}[xshift = 11cm, yshift = -4cm]
\node[align = right, anchor = east] at (0, 0.3) {for $j = 1, \ldots, m$\\ $\ell = 1, \ldots, 3$};
\end{scope}

\end{tikzpicture}
\caption{Edges related to the proof of Proposition~\ref{prop:segnp}.}
\label{fig:npproof}
\end{figure}

\emph{Step 2 (satisfiability implies zero cost solution):}
Assume that we have a truth assignment that satisfies the formula.
To show that there is a zero-cost rank assignment, we will construct
a change point function $\tau$. Then we show that solving \ctorprb with
this function results in a zero-cost rank assignment.
To define $\tau$, we first set
\[
\begin{split}
	\tau(p_i) = 1, \quad \tau(n_i) = 2,&
	\quad\text{if the $i$th variable is true, and} \\
	\tau(p_i) = 2, \quad \tau(n_i) = 1,&
	\quad\text{if the $i$th variable is false} \quad.
\end{split}
\]
Moreover, we set
$\tau(c_{j\ell}) = 2$ if the corresponding literal (taking possible negation into account) is true, 
and $\tau(c_{j\ell}) = 1$ otherwise.
Finally, we set
\[
	\tau(x_{j\ell}) = 1,\quad
	\tau(y_{j\ell}) = 2,\quad
	\tau(z_{j\ell}) = 2\quad.
\]

Let $H = G(\tau)$. We will show that $H$ is a DAG, which guarantees that the
rank assignment resulting from solving \ctorprb yields a zero cost. 
For every vertex in $v \in V(G)$, we will write $v^1$ and $v^2$ to refer
to the corresponding vertices in $H$, before and after the change point, respectively.

Assume there is a cycle $D$ in $H$.  Assume that the $i$th variable is set to true.
Then $p_i^1$ is a source and cannot be a part of $D$. Similarly, $n_i^2$ is a
sink. Since the only outgoing edge of $p^2_i$ goes to $n_i^2$, $p^2_i \notin
D$. Any $c_{j\ell}^1$ that corresponds to the \emph{negative} $i$th literal
is also a source. These vertices and $p_i^1$ are the only parents of $n_i^1$,
so $n_i^1 \notin D$. The case when the $i$th variable is false is symmetric.
In summary, $D$ does not contain
$p_i^1$,
$n_i^1$,
$p_i^2$, nor
$n_i^2$.
Since there are no other vertices joining vertices corresponding to different
clauses, $D$ must be among vertices $x_{j\ell}$, $y_{j\ell}$, $z_{j\ell}$,
$c_{j\ell}$ for a \emph{fixed} $j$.

Fix $j = 1, \ldots, m$.
$y_{j\ell}^2$ is a source and $z_{j\ell}^2$ is a sink, so they are outside of $D$.
Vertex
$x_{j\ell}^1$ is a sink, so it is outside of $D$.
The only outgoing edge of $y_{j\ell}^1$ goes to $x_{j\ell}^1$, so $y_{j\ell}^1 \notin D$.
The only outgoing edges of $z_{j\ell}^1$ go to $x_{j\ell}^1$ and $y_{j\ell}^1$, so $z_{j\ell}^1 \notin D$.
Vertex $c_{j\ell}^1$ is a sink within the subgraph corresponding to the clause,
so $c_{j\ell}^1 \notin D$.
In summary, $D$ may contain only $c_{j\ell}^2$, and $x_{j\ell}^2$.

The only possible cycle is then a $6$-cycle containing every $x_{j\ell}^2$
and $c_{j\ell}^2$. This is only possible if $\tau(c_{j\ell}) = 1$ for $\ell = 1, 2, 3$.
But this is a contradiction, since $j$th clause must be satisfied.
Consequently, $H$ is a DAG and solving \ctorprb for a given $\tau$ results in a rank assignment
that yields a zero cost.

\emph{Step 3 (zero cost solution implies satisfiability):}
Assume that $r$ is a rank assignment inducing a zero cost.
Let us write $\tau(v)$ to be the time stamp where the rank changes (if the rank is
constant, then set $\tau(v) = 1$). We can safely assume that $\tau(v) =1, 2$.
Define
$P = \set{i \mid \tau(p_i) = 1}$ and
$N = \set{i \mid \tau(n_i) = 1}$.
Note that $P \cap N = \emptyset$ and $P \cup N = [1, n]$,
otherwise the 1st row in Figure~\ref{fig:npproof} creates a cycle.
We set the variables whose indices are in $P$ to be true,
and the variables whose indices are in $N$ to be false.

Next, we prove that this assignment indeed solves \satprb.
We first claim that $\tau(x_{j\ell}) = 1$. Assume otherwise. Then
\[
\begin{split}
	r(y_{j\ell}; 0) & < r(x_{j\ell}; 0) = r(x_{j\ell}; 1) < r(y_{j\ell}; 1)\quad \text{and} \\
	r(z_{j\ell}; 0) & < r(x_{j\ell}; 0) = r(x_{j\ell}; 1) < r(z_{j\ell}; 1)\quad.
\end{split}
\]
Thus, $r(z_{j\ell}; 1) = r(z_{j\ell}; 2)$ and $r(y_{j\ell}; 1) = r(y_{j\ell}; 2)$.
But
\[
\begin{split}
	r(y_{j\ell}; 2) &  < r(x_{j\ell}; 2) < r(z_{j\ell}; 2) = r(z_{j\ell}; 1) 
	 < r(y_{j\ell}; 1) = r(y_{j\ell}; 2),
\end{split}
\]
which is a contradiction. Consequently, $r(x_{j\ell}; 0) \neq r(x_{j\ell}; 1)$, which forces $\tau(x_{j\ell}) = 1$.

Fix $j = 1, \ldots, m$.
Since $r(x_{j\ell}; 1) = r(x_{j\ell}; 2)$, then there is at least one $c_{j\ell'}$ 
for which $\tau(c_{j\ell'}) = 2$,
otherwise the 4th row in Figure~\ref{fig:npproof} creates a cycle.
Assume that $c_{j\ell'}$ corresponds to the positive $i$th literal.
This immediately implies that $\tau(p_i) = 1$,
otherwise the 2nd row in Figure~\ref{fig:npproof} creates a cycle.
By definition, $i \in P$, so the $i$th variable is true and the $j$th clause is satisfied.
Similarly, if $c_{j\ell'}$ corresponds to the negative $i$th literal,
then immediately $\tau(n_i) = 1$, so
$i \in N$ and the $i$th variable is false and the clause is satisfied.
Since this holds for every clause, \satprb is satisfied.
\qed
\end{IEEEproof}

\end{document}